\documentclass[aps,nofootinbib,preprint]{revtex4-1}

\usepackage{verbatim}
\usepackage[T1]{fontenc}
\usepackage[utf8]{inputenc}
\usepackage[american]{babel}
\usepackage{epsfig}
\usepackage{graphicx,subcaption,caption}
\usepackage{booktabs}
\usepackage{multirow}
\usepackage{dcolumn}
\usepackage{amsmath}
\usepackage{mathtools}
\usepackage{amsfonts}
\usepackage{amssymb}
\usepackage{ulem}
\usepackage{epstopdf}
\usepackage{bm}
\usepackage{siunitx}
\usepackage{braket}
\usepackage{enumitem}
\usepackage{soul}
\usepackage[table]{xcolor}
\usepackage{color}
\usepackage{transparent}
\usepackage{pifont}
\usepackage{enumitem}
\usepackage{CJKutf8}

\definecolor{navyblue}{rgb}{0.0, 0.0, 0.5}
\definecolor{royalblue}{rgb}{0.25, 0.41, 0.88}
\definecolor{cadmiumgreen}{rgb}{0.0, 0.42, 0.24}
\definecolor{blue-violet}{rgb}{0.54, 0.17, 0.89}
\definecolor{darkviolet}{rgb}{0.58, 0.0, 0.83}
\definecolor{orange(colorwheel)}{rgb}{1.0, 0.5, 0.0}

\usepackage{hyperref}
\hypersetup{
    colorlinks=true, % false: boxed links; true: colored links
    linkcolor=royalblue, % color of internal links (change box color with linkbordercolor)
    citecolor=magenta}

\usepackage{booktabs}
\usepackage{multirow}
\usepackage{dcolumn}
\usepackage{colortbl}

\begin{document}

\title{
Constraints on the simultaneous variation of the fine structure constant and electron mass
in light of DESI BAO data
}
\author{Yo Toda}
\email{y-toda@particle.sci.hokudai.ac.jp}
\affiliation{Department of Data \& Innovation, Kochi University of Technology, %Kita 10, Nisi 8, Kita-ku,
Tosayamada 782-8502, Japan \looseness=-1}
\affiliation{Department of Physics, Hokkaido University, %Kita 10, Nisi 8, Kita-ku,
Sapporo 060-0810, Japan \looseness=-1}

\author{Osamu Seto}
\email{seto@particle.sci.hokudai.ac.jp}
\affiliation{Department of Physics, Hokkaido University, %Kita 10, Nisi 8, Kita-ku,
Sapporo 060-0810, Japan \looseness=-1}

\begin{abstract}
%%%%%%%%%%%%%%%%%%%%%%
We study the cosmological constraints of the time variation of the electron mass $m_e$ and
the fine-structure constant $\alpha$, using data of cosmic microwave background, supernovae light curve and baryon acoustic oscillation (BAO) data including the recent DESI BAO DR2 measurements.
The results are slightly depending on the BAO data set included in the analysis.
The latest DESI BAO DR2 data strongly indicates that $m_e$ or $\alpha$ is slightly larger than the previous data from 6DF+SDSS and DESI BAO DR1.
We also compare the varying $m_e$ model, the varying $\alpha$ model, and the simultaneous variation of $m_e$ and $\alpha$.
When considering the Hubble tension, a larger electron mass is the most promising option and the variation of the fine-structure constants does not alleviate the tension.
%%%%%%%%%%%%%%%%%%%%%%
\end{abstract}
\preprint{EPHOU-25-005}

\maketitle

\section{Introduction}
\label{sec:introduction}

The standard cosmological model, known as the $\Lambda$CDM model, which includes cold dark matter (CDM) and a cosmological constant ($\Lambda$), is the most successful cosmological model in explaining the properties and evolution of our Universe.
The $\Lambda$CDM model consists of several fundamental physical constants, including the fine-structure constant $\alpha$ and the electron mass $m_e$, in addition to the cosmological constant.
These values are not necessarily fixed and might vary over time. The possibility of time-varying fundamental constants has been investigated by studying the Universe at different redshifts (see Refs.~\cite{Uzan:2002vq,Uzan:2010pm,Martins:2017yxk,Uzan:2024ded} for a review).
Using observational data of Big Bang Nucleosynthesis (BBN)~\cite{Cooke:2017cwo,Aver:2015iza}, we obtain a constraint on the fine-structure constant during the BBN era:
$-1.2\% < \Delta \alpha/\alpha < 0.4\%$ (68\% C.L.)~\cite{Seto:2023yal}.
From quasar spectra observations~\cite{Murphy:2016yqp, Evans:2014yva, Murphy:2017xaz, Songaila:2014fza}, they derive a constraint on the fine-structure constant at lower redshifts ($0<z<3$), with an upper bound $|\Delta \alpha/\alpha| \lesssim 10^{-5}$.

Observations of the Cosmic Microwave Background (CMB) have also played a crucial role in constraining time-varying fundamental constants.
In models with a varying fine-structure constant or electron mass, the Thomson scattering cross-section $\sigma_T$  and the time of last scattering $t_*$ are modified.
Taking these effects into account, Planck provided constraints on the time variation of these constants at redshift $z \simeq 10^3$~\cite{Planck:2014ylh}.
As observational precision has improved, the Hubble tension—a discrepancy between the present Hubble parameter ($H_0$) measured from local observations~\cite{Riess:2021jrx,Riess:2023bfx,Riess:2024ohe,Wong:2019kwg,Freedman:2019jwv,Freedman:2020dne,Freedman:2021ahq} and its inference from early cosmological observations~\cite{Planck:2018vyg, Beutler:2011hx,Ross:2014qpa,BOSS:2016wmc, Schoneberg:2019wmt}—has become more concrete.
Among the various proposed solutions, the varying electron mass model is considered one of the most promising approaches to alleviating the Hubble tension~\cite{Schoneberg:2021qvd}.
This model has been extensively studied from multiple perspectives~\cite{Hoshiya:2022ady, Seto:2022xgx, Seto:2024cgo, Toda:2024ncp, Toda:2024uff, Sekiguchi:2020teg, Sekiguchi:2020igz, Hart:2017ndk, Hart:2019dxi, Chluba:2023xqj, Smith:2018rnu, Lynch:2024hzh, Schoneberg:2024ynd, Khalife:2023qbu}.
 
The Dark Energy Spectroscopic Instrument (DESI) survey has released the results of its most recent BAO measurements from DR1~\cite{DESI:2024uvr,DESI:2024lzq, DESI:2024mwx} and DR2~\cite{DESI:2025zpo,DESI:2025zgx}.
These results indicate that that the standard cosmological constant dark energy model, which is characterized by $(w, w_a) = (-1, 0)$, has a discrepancy of $2.8\sigma$–$4.2\sigma$  when combined with DESI BAO, CMB, and SNe data~\cite{DESI:2025zgx}. Here, $w_a$ is the change in the equation-of-state parameter $w$ with respect to the scale factor $a$. 
It was found in a previous study~\cite{Seto:2024cgo} that the surprising discrepancy in the DESI BAO results may be attributed to a shorter sound horizon during recombination, which could be a result of an increase in electron  mass\footnote{A shorter sound horizon can also be achieved through early dark energy~\cite{Poulin:2018dzj,Poulin:2018cxd,Braglia:2020bym,Agrawal:2019lmo,Ye:2020btb,Smith:2019ihp,Lin:2019qug,Niedermann:2019olb,Niedermann:2020dwg, Seto:2021xua}, which increases the energy density in the early universe and consequently reduces the sound horizon.}. Other implications of recent DESI results on dark energy and neutrinos
 have been studied for example in Refs.~\cite{Tada:2024znt, Wang:2024hks, Giare:2024smz, Giare:2024gpk, Luongo:2024fww, Colgain:2024xqj, Jiang:2024viw}.

In this paper, we obtain more stringent constraints on time varying $m_e$ from updated the DESI BAO and Planck CMB data. Furthermore, we explore the possibility of changes in the fine-structure constant, as it would naturally vary when considering a fundamental theory where the electron mass can vary due to the dynamics of a dilation field or extra-dimensional space~\cite{Carroll:1998zi,Chiba:2006xx}.

This paper is organized as follows.  We present the explanation of the model in Sec~\ref{sec:models}, the method and the datasets of our analysis in Sec.~\ref{sec:data}, the results in Sec.~\ref{sec:results}, and the summary in Sec~\ref{sec:conclusions}.

\section{Models}
\label{sec:models}

In this paper, we focus on the role of electrons as well as the varying fine-structure constant during the recombination epoch.  

\subsection{Varying electron mass}

The electron mass in the early Universe is thought to have been different from its present value ($m_{e0}=511 \,\mathrm{keV}$) in the varying electron mass model, which transitions to its present value after recombination.
The increase in electron mass results in an increase in hydrogen energy $E$ levels and the energy (or corresponding frequency $\nu$) of Lyman-alpha photons, which both scale proportionally with the electron mass as
\begin{subequations}
\begin{align}
E & \propto m_{e}, \\
\nu & \propto m_{e}.
\end{align}
\label{eq:Eme}
\end{subequations}
If the electron mass is larger, hydrogen energy levels increase and the frequency of photons needed to excite a hydrogen also increases. 
The photon frequency is inversely proportional to the cosmic scale factor as $\nu\propto 1/a$, resulting in photons losing the necessary energy to excite hydrogen atoms at a earlier time.
As a result, a larger electron mass leads to earlier recombination and a shorter sound horizon. 
To keep same angular scale of the acoustic peaks in the CMB power spectrum have as observed, a shorter sound horizon must have a smaller angular diameter distance, which can infer larger $H_0$ values.

If the electron mass is larger, the Silk dampling scale defined by
\begin{equation}
\lambda_D^2=\frac{1}{6}\int^{\eta_\mathrm{dec}}_{0}\frac{d\eta}{\sigma_Tn_ea}\left[\frac{R^2+\frac{16}{15}(1+R)}{(1+R)^2}\right],
\end{equation}
seems to insrease, because the Thomson scattering cross section $\sigma_{T}$
\begin{equation}
\sigma_{T} \propto \frac{1}{m_e^2},
\label{eq:sigmame}
\end{equation}
 scales inversely the squared electron mass.
Here, $n_e$ is the number density of free electrons, $R=3\rho_b/4\rho_r$ is the baryon-radiation ratio, and $\eta_\mathrm{dec}$ is the conformal time of decoupling. 
The larger the electron mass, the earlier the decoupling time $\eta_\mathrm{dec}$ as previously mentioned.
Due to compensating for these two effects, the variable electron mass model does not significantly affect the silk damping length and alleviates Hubble tension without affecting the CMB fitting.

The top panel of Fig.~\ref{fig:TT} displays the CMB TT angular power spectra for different values of $m_e/m_{e,0}$ for fixed other cosmological parameters. 
The position of each acoustic peak is moved to a higher multipole $\ell$ by a larger electron mass, which indicates earlier recombination, but it does not significantly increase the peak heights beyond the first and second peaks. In this figure, we also include the other minor contributions: the photoionization cross sections, the recombination coefficients, the ionization coefficients, $K$ factors, Einstein $A$ coefficients, and the two-photon decay rates. The detail of those effects can be found in Ref.~\cite{Planck:2014ylh}.

As previously mentioned, the varying electron mass model provides a feasible solution to the Hubble tension without significantly compromising the fit to the CMB data. 
This can be understood as a result of parameter degeneracy between cold dark matter density $\omega_\mathrm{c}=\Omega_\mathrm{c}h^2$, baryon density $\omega_\mathrm{b}=\Omega_\mathrm{b} h^2$, 
 and the electron mass $m_{e}/m_{e,0}$, which allows a larger value of Hubble constant $H_{0}$ as discussed in Ref.~\cite{Sekiguchi:2020teg}. When BAO measurements are taken into account, this degeneracy can be resolved.
A larger electron mass infering a larger $H_0$ is realized for larger $\omega_\mathrm{c}$ and $\omega_\mathrm{b}$, which also inferes larger $\theta_d(z)\equiv r_d/D_M$, where $r_d=\int_{z_d}^{\infty}\frac{c_s(z)}{H(z)}dz$ is the sound horizon, $D_M(z)=\frac{c}{H_0} \int_0^z\frac{dz'}{H(z')/H_0}$ is the diameter distance, and $z_d$ is the redshift of the drag epoch.
The analysis using the DESI BAO DR1 data suggests a possibility for a $1\%$ larger electron mass~\cite{Seto:2024cgo} despite the latest SDSS BAO data~\cite{eBOSS:2020yzd} not supporting a larger electron mass of $m_e/m_{e,0}$~\cite{Khalife:2023qbu}.

\subsection{Varying fine-structure constant}
 
\begin{figure}[ht]
\includegraphics[width=17cm]{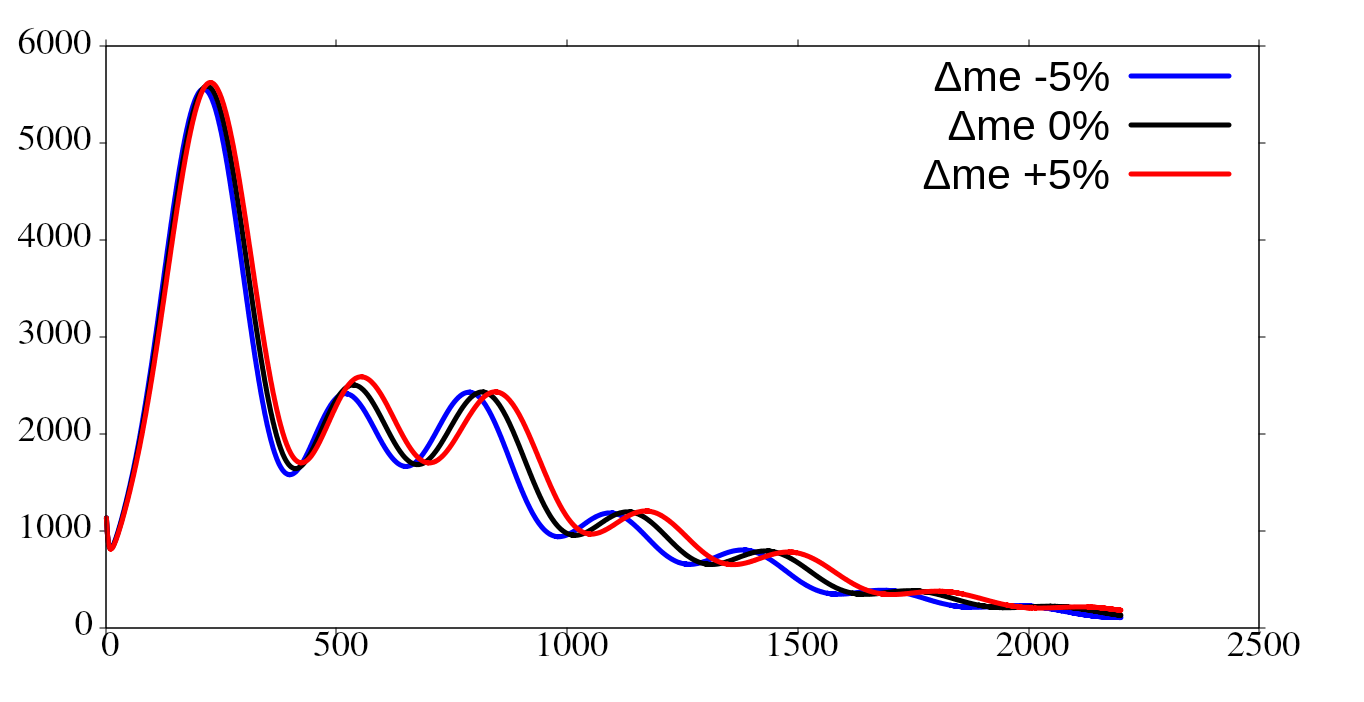}
\includegraphics[width=17cm]{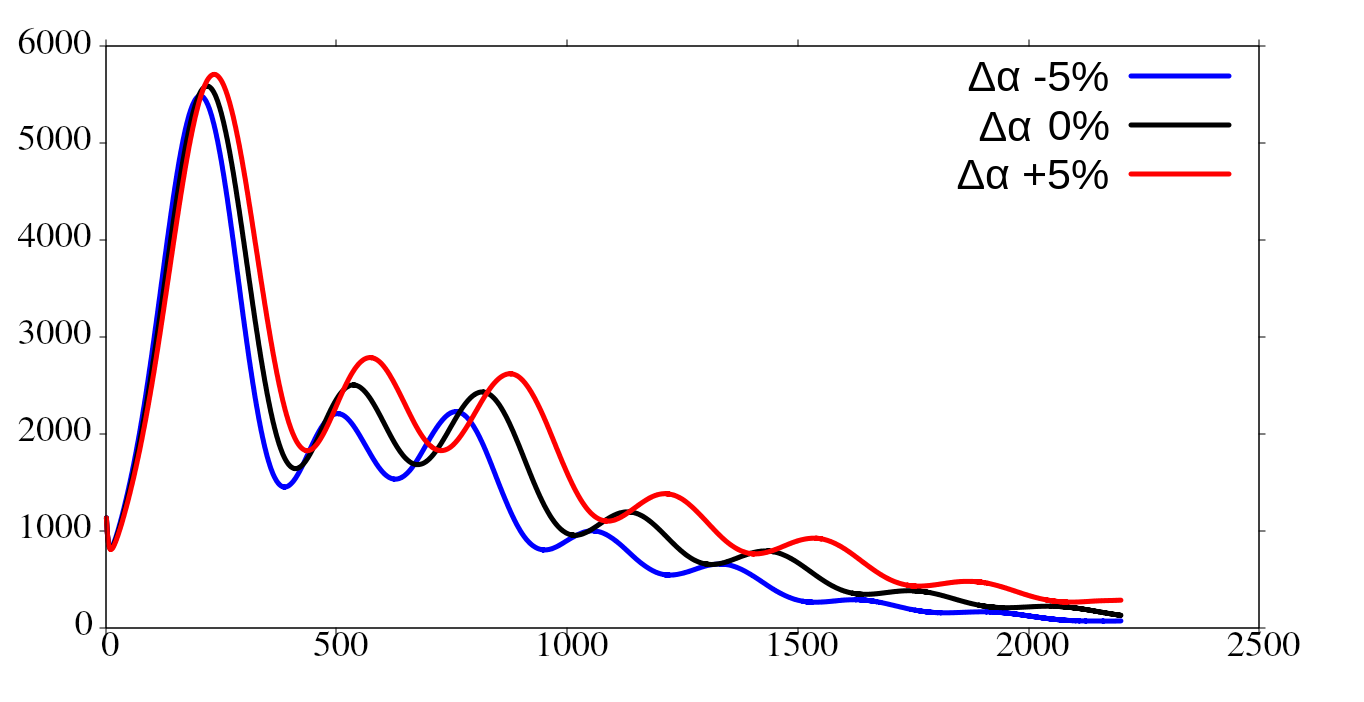}
\centering
\caption{CMB TT angular power spectra for different values of $m_e$ (top) and $\alpha$ (bottom). The blue, black, red curves stand for variations of $-5\%$, $0\%$, and $+5\%$, respectively.}
\label{fig:TT} 
\end{figure}

The dependence of the energy level of hydrogen and the Thomson scattering cross section 
on the fine structure constant as well as the electron mass is expressed as
\begin{align}
& E\propto m_e \alpha^2,  \label{eq:E_to_me_alpha} \\ 
& \sigma_{T} \propto \frac{\alpha^2}{m_e^2}.
\end{align}
The fine structure constant increase leads to an earlier recombination, which is the same effect as increasing $\sqrt{m_e}$.
On the other hand, a larger fine structure constant increases the Thomson scattering cross section, which is equivalent to the decrease of electron mass and induces further reduction of the Silk damping scale in addition to it by the earlier recombination. 
A larger fine-structure constant leads to a larger amplitude of the acoustic peaks, as the damping is less effective.
The bottom panel of Fig.~\ref{fig:TT} displays the CMB TT angular power spectra for different values of $\alpha/\alpha_0$ with $\alpha_0=1/137$. It is seen that a larger fine-structure constant shifts the peak positions to higher multipoles and increases their amplitudes.

\section{datasets and methodology}
\label{sec:data}

We perform a MCMC analysis of the time-varying
electron mass model and the time-varying fine-structure constant model, using the public MCMC code \texttt{COBAYA}\footnote{\texttt{COBAYA} and the cosmological data are available at \url{https://github.com/CobayaSampler.}}~\cite{Torrado:2020dgo},
requiring the convergence $R-1<0.02$.
We also use the cosmological Boltzmann code \texttt{CAMB}~\cite{Lewis:1999bs,Howlett:2012mh}
and recombination code \texttt{recfast}~\cite{Scott:2009sz} above modifications implemented.
We use the following default data sets. The first two are refered to as $\mathcal{D}$ always included in our analysis, and we examine difference in infered cosmological parameters for several BAO data:
\begin{itemize}
\item The temperature and polarization likelihoods for high $l$ \textsc{CamSpec} which use \texttt{NPIPE} (Planck PR4) data~\cite{Planck:2019nip,Rosenberg:2022sdy}.
We also include low$l$ \texttt{Commander}, lowE \texttt{SimAll}~\cite{Aghanim:2019ame}, and CMB lensing~\cite{Planck:2018lbu}.
\item Light curve of SNeIa  from \texttt{Pantheon+}~\cite{Pan-STARRS1:2017jku}.
\item BAO from 6dF~\cite{Beutler:2012px}, and SDSS DR7~\cite{Ross:2014qpa, Alam:2020sor}. We use the distance and growth rate data, called 6DF+SDSS.
\item DESI BAO DR1~\cite{DESI:2024mwx}.  We use the distance data called DESI BAO-DR1.
\item DESI BAO DR2~\cite{DESI:2025zgx}.  We use the distance data called DESI BAO-DR2.
%\item Local $H_0$ measurement from \cite{Riess:2020fzl}, in terms of magnitude measurement.
%\item Combination of galaxy clustering and weak lensing data from the first year of the
%Dark Energy Survey (DES Y1) \cite{Abbott:2017wau}.
\end{itemize}

\section{Results}
\label{sec:results}
%We present the results of models we have examined. 
%The 1D and 2D marginalized posteriors of different cosmological parameters are shown in Fig.~\ref{fig:me}, \ref{fig:alpha}, \ref{fig:Comparison} and the 68\% confidence level constraints are listed in Table~\ref{table}.

\begin{figure}[ht]
\includegraphics[width=17cm]{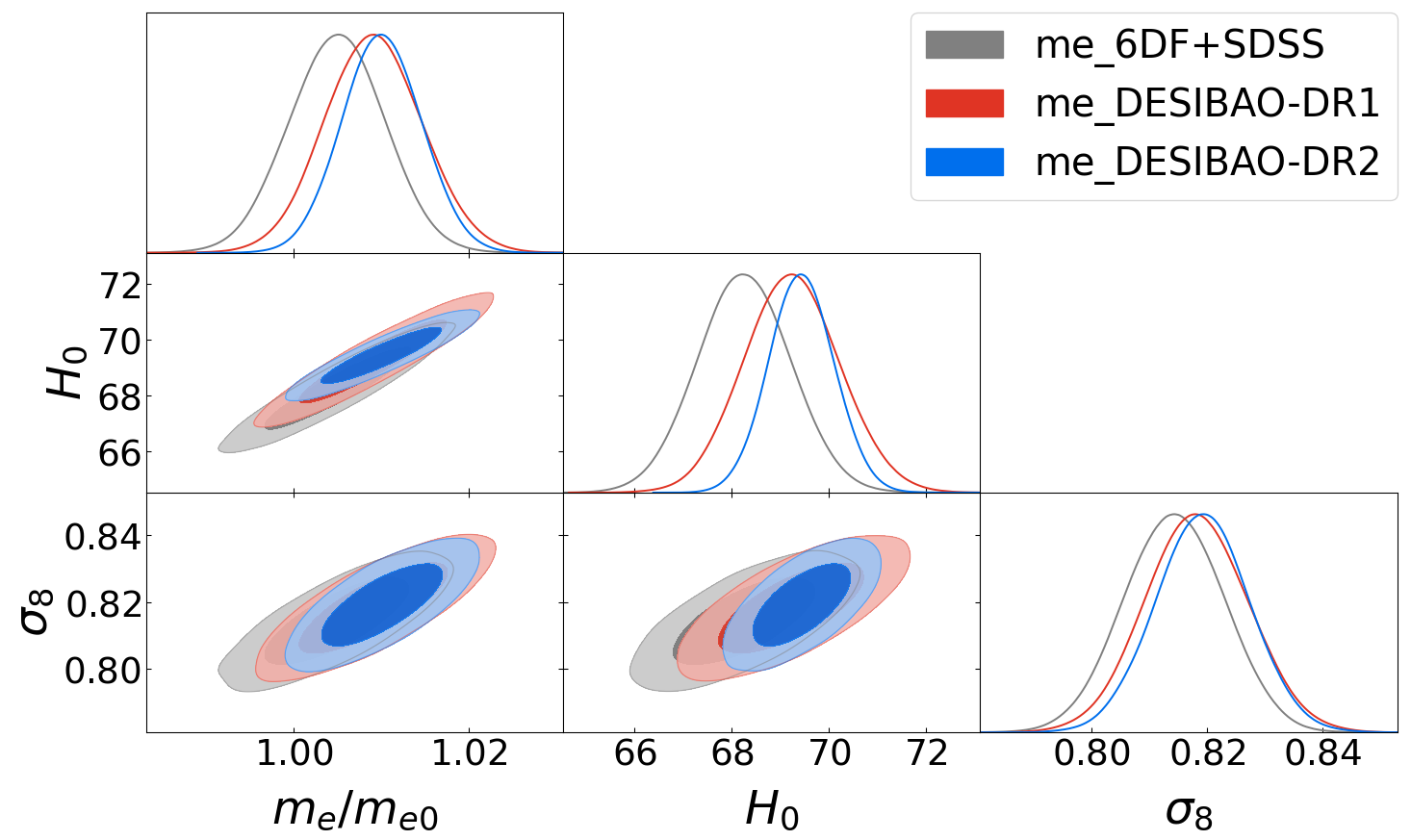} 
\centering
\caption{Posterior distributions of some parameters for the varying $m_e$ model. 
Different colors of grayish, reddish and bluish contours stand for different datasets descriebd in key. }
\label{fig:me} 
\end{figure}

%%%%%%%%%%%%%%%me
We show the 1D and 2D marginalized posteriors of the varying electron mass model for the datasets: $\mathcal{D}$+6DF+SDSS, $\mathcal{D}$+DESI BAO DR1, and $\mathcal{D}$+DESI BAO DR2 in Fig.~\ref{fig:me}.
The posteriors from $\mathcal{D}$+DESI BAO DR2 are consistent with those from the other datasets.
We find that a slightly larger electron mass $m_e/m_{e,0} = 1.0101\pm0.0046$ ($\mathcal{D}$+DESI BAO DR2) is preferred and standard value $m_e/m_{e,0}=1$ locates at more than $2\sigma$ away.
The analysis using DESI BAO DR1 data also favors a $1\%$ larger electron mass with larger uncertainty than that with DESI BAO DR2 data.
%%%%%%%%
DESI BAO results shows that the sound horizon relative to the diameter distance $r_d/D$ is a bit longer than that in the standard $\Lambda$CDM as shown in Fig.~6 in the DESI paper~\cite{DESI:2025zgx}. 
The measured longer $r_d/D$ indicates a larger product of the sound horizon and the Hubble constant $r_d H_0$, because the diameter distance $D$ is inversely proportional to the Hubble constant $H_0$.
%%%%%%%
The vaying electron mass model with a larger electron mass shorten the sound horizon $r_d$ with inferring larger Hubble constant $H_0$ and increase the product $r_dH_0$ slightly\footnote{See Ref.~\cite{Seto:2024cgo} for the detailed discussion.}, as the data including the DESI BAO data indicate~\cite{DESI:2025zgx}.
%%%%%%%%
As a result, DESI BAO data seem to support a $1\%$ larger electron mass as shown in Fig.~\ref{fig:me}.

\begin{figure}[ht]
\includegraphics[width=16cm]{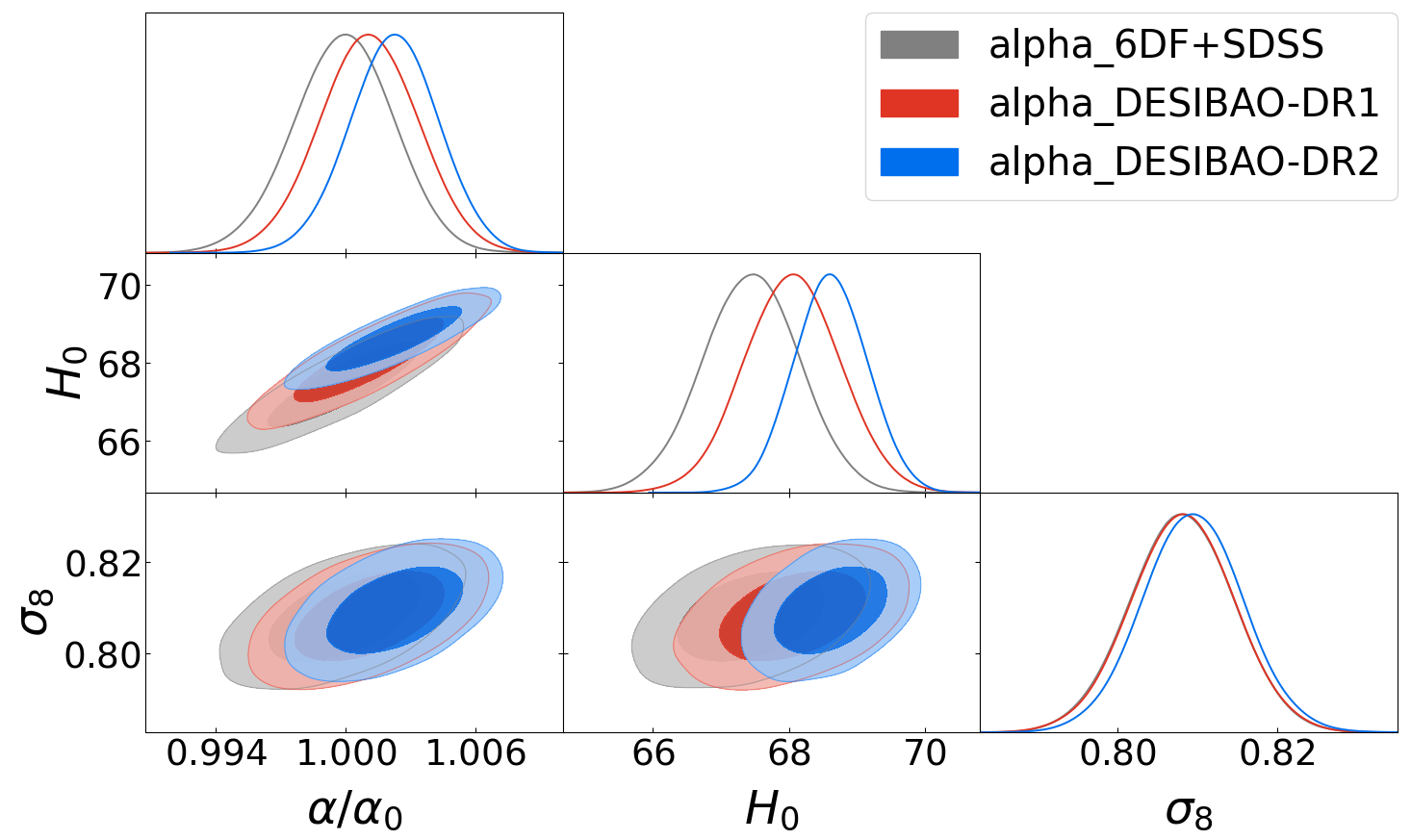} \caption{ Posterior distributions of some parameters for varying $\alpha$
model. Different colors of grayish, reddish and bluish contours stand for different datasets. }
\label{fig:alpha} 
\end{figure}

%%%%%%%%%%%%%%%%%%%Alpha
Next, we show the 1D and 2D marginalized posteriors of the varying fine-structure constant model for the datasets: $\mathcal{D}$ + 6DF + SDSS, $\mathcal{D}$ + DESI BAO DR1, and $\mathcal{D}$ + DESI BAO DR2 in Fig.~\ref{fig:alpha}.
Similar to the varying electron mass case, we find that a slightly larger fine-structure constant ($\alpha/\alpha_0 = 1.0023\pm0.0021$ ($\mathcal{D}$+DESI BAO DR2)) is preferred from DESI BAO DR2, although the significance is only at the $1\sigma$ confidence level.

\begin{figure}[ht]
\includegraphics[width=17cm]{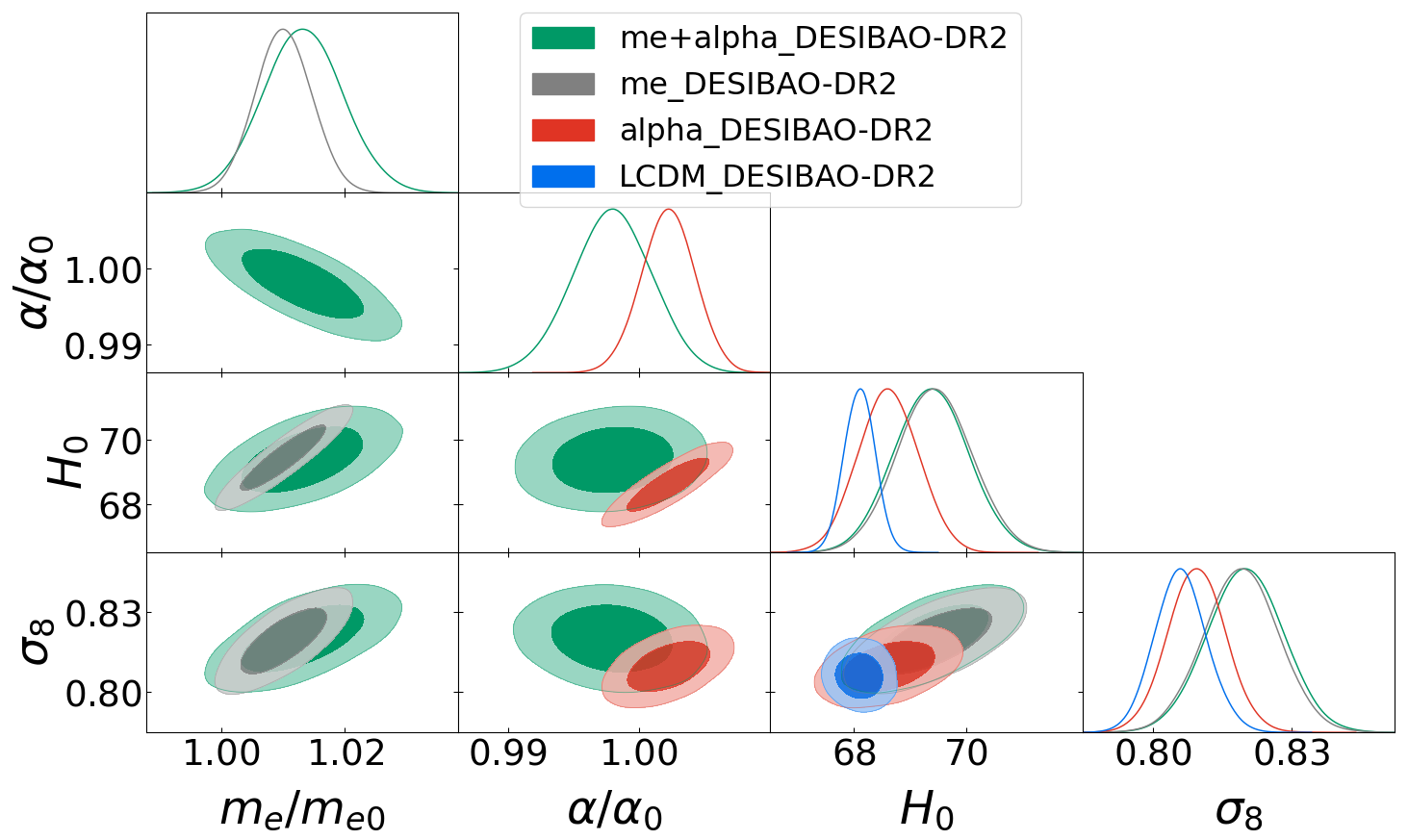} 
\centering
\caption{Posterior distributions of some parameters for the varying $m_e$ model. 
Different colors of greenish, grayish, reddish and bluish contours stand for different models. }
\label{fig:Comparison} 
\end{figure}

%%%%%%%%%%%%%%%%%%%comparison
For comparison of models, we show the 1D and 2D marginalized posteriors of each models of $\Lambda$CDM (bluish), varying $m_e$ (grayish), varing $\alpha$ (reddish) and varying both $m_e$ and $\alpha$ (greenish)
for the common datasets: $\mathcal{D}$ + DESI BAO DR2 in Fig.~\ref{fig:Comparison}.
The green contours show that there are correlations, in other words parameter degeneracy,
between $m_e/m_{e,0}$ and $\alpha/\alpha_0$. 
Along this degeneracy, an increase in the electron mass can be compensated by a decrease in the fine-structure constant to reproduce the same recombination time, which can be understood through Eq.~(\ref{eq:E_to_me_alpha}).

Finally, in Table.~\ref{table}, we list the $68\%$ confidence level constraints of some cosmological parameters and those Gaussian tension. The Gaussian tensions for $H_{0}$ and $\sigma_{8}$ are calculated as
\begin{equation}
T_{H_{0}}=\frac{H_{0~\mathrm{\mathcal{D}+DESI-DR2}}-73.30}{\sqrt{\sigma_{\mathrm{\mathcal{D}+DESI-DR2}}^{2}+1.04^{2}}}\,,
\end{equation}
for the Hubble tension with the SHOSE measurement~\cite{Riess:2021jrx} and
\begin{equation}
T_{\sigma_{8}}=\frac{\sigma_{8~\mathrm{\mathcal{D}+DESI-DR2}}-0.802}{\sqrt{\sigma_{\mathrm{\mathcal{D}+DESI-DR2}}^{2}+\frac{0.022^2+0.018^2}{2}}}\,,
\end{equation}
for the $\sigma_{8}$ tension with Kilo-Degree Survey (KiDS-Legacy)~\cite{Stolzner:2025htz}, respectively. 
Since the latest value $\sigma_8=0.802^{+0.022}_{-0.018}$ from KiDS-Legacy
 is a rather optimistic value between various measurements of $\sigma_8$,
 it relaxes the $\sigma_8$ tension.

In Tab.~\ref{table}, from the viewpoint of the Hubble tension, we find that the varying electron mass model is the most effective solution among the four models, reducing the tension to $3.12\sigma$, which is $1.7\sigma$ lower than that of the $\Lambda$CDM model. 
No additional improvement in alleviating the Hubble tension is shown when both the electron mass and the fine-structure constant are varied simultaneously.
Regarding the $\sigma_8$ tension, although the values are slightly worse in varying electron mass model, the tensions are below $1\sigma$ for all models, since we have compared with the latest KiDS-Legacy value $\sigma_8=0.802$.

\begin{table*}
\[
\begin{tabular}{lccccc|}
\hline \hline
 Parameter  & \ensuremath{\Lambda}CDM & varying\,\ensuremath{m_{e}} & varying\,\ensuremath{\alpha} &  varying\,\ensuremath{m_{e}}+\ensuremath{\alpha}\\
\hline  {\boldmath\ensuremath{m_{e}/m_{e0}}}  & - & \ensuremath{1.0101\pm0.0046} & - &  \ensuremath{1.0133\pm0.0065}\\
{\boldmath\ensuremath{\alpha/\alpha_{0}}}  & - & - & \ensuremath{1.0023\pm0.0021} &  \ensuremath{0.9980\pm0.0030}\\
\ensuremath{H_{0}}[km/s/Mpc]  & \ensuremath{68.11\pm0.28} & \ensuremath{69.44\pm0.67} & \ensuremath{68.61\pm0.54} &  \ensuremath{69.38\pm0.67}\\
\ensuremath{\sigma_{8}}  & \ensuremath{0.8060\pm0.0057} & \ensuremath{0.8192\pm0.0081} & \ensuremath{0.8095\pm0.0063} &  \ensuremath{0.8201\pm0.0083}\\
\hline 
$T_{H_0}$ & 4.82$\sigma$ & 3.12$\sigma$& 4.00$\sigma$& 3.16$\sigma$ \\
$T_{\sigma_8}$ &    0.19$\sigma$&     0.79$\sigma$&   0.36$\sigma$& 0.83$\sigma$\\
\hline \hline
\end{tabular}
\]
\caption{68\% constraints and Gaussian tension to other measurements from the dataset $\mathcal{D}$+DESI BAO DR2.
\label{table}  }
\end{table*}

\section{Conclusions}
\label{sec:conclusions} 

The varying electron mass model and the varying fine-structure constant model are investigated in this paper, with the recent BAO measurements being taken into account.
In the varying electron mass model, we find that the larger electron mass is preferred as:
\begin{align}
&
\begin{cases}
m_{e}/m_{e0}=1.0049\pm 0.0055  \\
H_{0}=68.26\pm 0.96  \,\mathrm{km}/\mathrm{s}/\mathrm{Mpc}\\
\end{cases}(\mathrm{CMB+SNeIA+BAO(6DF+SDSS)}), \\
&
\begin{cases}
m_{e}/m_{e0}=1.0101\pm0.0046\\
H_{0}=69.44\pm 0.67 \,\mathrm{km}/\mathrm{s}/\mathrm{Mpc}  \\
\end{cases}(\mathrm{CMB+SNe\,Ia+BAO(DESI\,DR2)}).
\end{align}
In the analysis using the conventional BAO (6DF+SDSS), the standard value $m_e/m_{e,0}=1$ is in 68\% C.L., whereas,
in the analysis using the DESIl BAO (DR2), the standard value $m_e/m_{e,0}=1$ is out more than $2\sigma$,

In the varying fine-structure constant model, we have the following parameter constraints:
\begin{align}
&
\begin{cases}
\alpha/\alpha_{0}=0.9999\pm 0.0023  \\
H_{0}=67.44\pm 0.72  \,\mathrm{km}/\mathrm{s}/\mathrm{Mpc}\\
\end{cases}(\mathrm{CMB+SNe\,Ia+BAO(6DF+SDSS)}), \\
&
\begin{cases}
\alpha/\alpha_0=1.0023\pm0.0021\\
H_{0}=68.61\pm 0.54 \,\mathrm{km}/\mathrm{s}/\mathrm{Mpc}  \\
\end{cases}(\mathrm{CMB+SNe\,Ia+BAO(DESI\,DR2)}).
\end{align}
We find that a slightly larger fine-structure constant is preferred for DESI BAO DR2 data. This result indicates that Hubble tension solutions with earlier recombination and shorter sound horizon are favored when we take account of DESI BAO measurement data.While it is preferable to have a larger electron mass, it is not preferable to have a significantly different value of fine-structure constant in the early Universe from the present value.
This is because large variations in the fine-structure constant is not allowed by its effect to the Silk damping length.

While the variation of $\alpha$ alone is not not preferred, there are two parameter degenerates that are approximately given by $\Delta \alpha \simeq -\frac{1}{2}\Delta m_e$. This feature might be beneficial in constructing particle physics models that incorporate variations in fundamental constants.

\begin{acknowledgments}
\noindent This work was supported by JSPS KAKENHI Grant No. 23K03402
(O. S.), and JST SPRING, Grant No. JPMJSP2119 (Y. T.). 
\end{acknowledgments}

\bibliography{me_DESI}

\end{document}